\documentclass[prd,amsfonts,onecolumn,superscriptaddress,aps,nofootinbib,11pt]{revtex4}

\pdfoutput=1
\usepackage[utf8]{inputenc}
\usepackage{amsmath,amssymb,mathrsfs}
\usepackage{graphicx}
\usepackage{enumerate}
\usepackage{wrapfig}
\usepackage{multirow}
\usepackage{array}
\usepackage[usenames,dvipsnames]{xcolor} 
\usepackage{hyperref}
\hypersetup{colorlinks=true,
		breaklinks=true,
    linkcolor=Blue,          
    citecolor=Plum,        
    urlcolor=YellowOrange           
}
\def\be{\begin{equation}}
\def\ee{\end{equation}}
\def\bea{\begin{eqnarray}}
\def\eea{\end{eqnarray}}

\def\3311{$\mathrm{SU(3) \otimes SU(3)_L \otimes U(1)_X \otimes U(1)_{N}}$ }
\def\vev#1{\left\langle #1\right\rangle}

\newcommand{\sm}{{Standard Model }}

\providecommand{\be}{ \begin{equation} } 
\providecommand{\ee}{ \end{equation} }
\providecommand{\bea}{\begin{eqnarray}}
\providecommand{\eea}{\end{eqnarray}}
\providecommand{\nn}{\nonumber}

\usepackage{soul}

\begin{document}
\title{Scotogenic neutrino masses and dark matter stability\\
from residual gauge symmetry}

\author{Julio Leite\footnote{Talk at NDM 2020, Hurgada, Egypt, January
2020.}}%
\affiliation{AHEP Group, Institut de F\'{i}sica Corpuscular --
  C.S.I.C./Universitat de Val\`{e}ncia, Parc Cient\'ific de Paterna.\\
 C/ Catedr\'atico Jos\'e Beltr\'an, 2 E-46980 Paterna (Valencia) - SPAIN}
\affiliation{Centro de Ci\^encias Naturais e Humanas, Universidade Federal do ABC, Santo Andr\'e-SP, Brasil}

\author{Oleg Popov}%
\affiliation{Institute of Convergence Fundamental Studies, Seoul National University of Science and Technology,\\ Seoul 139-743, Republic of Korea}
\affiliation{Department of Physics, Korea Advanced Institute of Science and Technology, \\291 Daehak-ro, Yuseong-gu, Daejeon 34141, Republic of Korea}

\author{Rahul Srivastava}%
\affiliation{AHEP Group, Institut de F\'{i}sica Corpuscular --
  C.S.I.C./Universitat de Val\`{e}ncia, Parc Cient\'ific de Paterna.\\
 C/ Catedr\'atico Jos\'e Beltr\'an, 2 E-46980 Paterna (Valencia) - SPAIN}
\affiliation{ India Institute of Science Education and Research - Bhopal, Bhopal Bypass Road, Bhauri, 462066, Bhopal, India}

\author{Jos\'{e} W. F. Valle}%
\affiliation{AHEP Group, Institut de F\'{i}sica Corpuscular --
  C.S.I.C./Universitat de Val\`{e}ncia, Parc Cient\'ific de Paterna.\\
 C/ Catedr\'atico Jos\'e Beltr\'an, 2 E-46980 Paterna (Valencia) - SPAIN}

\begin{abstract}
In the context of the $\mathrm{SU(3)_c \otimes SU(3)_L \otimes U(1)_X \otimes U(1)_{N}}$ (3-3-1-1) extension of the standard model, we show how the spontaneous breaking of the gauge symmetry gives rise to a residual symmetry which accounts for dark matter stability and small neutrino masses in a scotogenic fashion.
As a special feature, the gauge structure implies that one of the light neutrinos is massless and, as a result, there is a lower bound for the $0\nu\beta\beta$ decay rate.
\end{abstract}

\maketitle
\section{Introduction}
\label{Sect:intro}

The lack of a viable dark matter (DM) candidate within the \sm is one of the most pressing issues requiring new physics. In addition to new particles, the existence of DM requires new symmetries to stabilise the corresponding candidate on cosmological scales, {\it e.g.} R-parity symmetry in supersymmetric schemes~\cite{Jungman:1995df}. Another important open question in need of new physics is that of neutrino masses, which are necessary to account for neutrino oscillation data \cite{deSalas:2017kay}.

In order to deal simultaneously with the DM and neutrino mass issues, ``low-scale'' realisations where dark matter appears as a radiative mediator of neutrino mass generation, known as scotogenic models, are appealing alternatives. 
In its original version~\cite{Ma:2006km}, the symmetry stabilising dark matter is also responsible for the radiative origin of neutrino masses in a very elegant way. 
Yet, in this case as well as in other proposals \cite{Hirsch:2013ola,Merle:2016scw}, the stabilisation symmetry is introduced in an \textit{ad hoc} manner. 

Extending the \sm gauge symmetry can provide a natural setting for a theory of dark matter where stabilisation is automatic~\cite{Alves:2016fqe,Dong:2017zxo,Kang:2019sab}.
Such electroweak extensions involve the SU(3)$_{\rm L}$ gauge symmetry, which remarkably explains the observed number of fermion families from the anomaly cancellation requirement~\cite{Singer:1980sw,Pisano:1991ee,Frampton:1992wt}. 
In the extended models discussed in~\cite{Alves:2016fqe,Dong:2017zxo,Kang:2019sab}, DM stability follows naturally from the spontaneous breaking of the extended gauge symmetry into the residual matter-parity symmetry, $M_P$, a non-supersymmetric version of R-parity. 

We show here how scotogenic neutrino masses and automatic DM stabilisation are intrinsically linked in a 3-3-1-1 model \cite{Leite:2019grf}. The present construction has the special feature of predicting that one of the light neutrinos is massless, which, in turn, implies that there is a lower bound for neutrinoless double beta decay rate. This feature arises in a novel way when compared to other schemes in the literature, such as the ``incomplete'' seesaw mechanism~\cite{Schechter:1980gr} or similar radiative mechanisms~\cite{Reig:2018ztc}.

\section{The model}
\label{sec:model}


In our model \cite{Leite:2019grf}, the electric charge and $B-L$ generators are given by $Q  =  T_3 - (1/\sqrt{3}) T_8 + X~$ and $B-L  =  -(2/\sqrt{3}) T_8 + N~$, respectively, where $T_3$ and $T_8$ are the diagonal $\mathrm{SU(3)_L}$ generators, while $X$ and $N$ are the $\mathrm{U(1)_X}$ and $\mathrm{U(1)_N}$ generators, respectively. 
Notice that, due to the extra $U(1)_N$ symmetry, the $B-L$ symmetry is fully gauged. 
The lepton and scalar fields and their respective transformations are summarised in Table \ref{tab}. In the lepton sector, the left-handed fields appear as $SU(3)_L$ triplets, while the right-handed are $SU(3)_L$ singlets. The scalar sector contains five triplets and one singlet. 
\begin{table}[h!]
    \centering
    \begin{tabular}{|c|c|c|c|c|c|c|}
        \hline
\hspace{0.2cm} Field \hspace{0.2cm}&\hspace{0.2cm} SU(3)$_L$ \hspace{0.2cm}& \hspace{0.2cm}U(1)$_X$ \hspace{0.2cm}& \hspace{0.2cm}U(1)$_N$  \hspace{0.2cm}&\hspace{0.2cm} $Q$\hspace{0.2cm} &\hspace{0.2cm} $B-L$\hspace{0.2cm} &\hspace{0.2cm} $M_P = (-1)^{3(B-L)+2s}$ \hspace{0.2cm}\\
\hline\hline
        $l_{aL}$ & {\bf 3} & $-1/3$ & $-2/3$ & $(0,-1,0)^T$ & $(-1,-1,0)^T$ & $(++-)^T$ \\
       $ e_{a R}$ & {\bf 1} & $-1$ & $-1$ & $-1$ & $-1$ & $+$ \\
       \hline
       $ \nu_{i R}$ & {\bf 1} & $0$ & $-4$ & $0$& $-4$&$-$ \\
       $  \nu_{3R}$ & {\bf 1} & $0$ & $5$ & $0$& $5$&$+$ \\ $  N_{aR}$ & {\bf 1} & $0$ & $0$ & $0$& $0$& $-$  \\
       \hline\hline
        $\eta$ &  {\bf 3} & $-1/3$ & $1/3$ &  $(0,-1,0)^T$&  $(0,0,1)^T$&$(++-)^T$ \\
        $\rho$ & {\bf 3} & $2/3$ & $1/3$ &  $(1,0,1)^T$&  $(0,0,1)^T$&$(++-)^T$ \\
        $\chi$ & {\bf 3} & $-1/3$ & $-2/3$ &  $(0,-1,0)^T$&  $(-1,-1,0)^T$&$(--+)^T$ \\
        $\sigma$ & {\bf 1} & $0$ & $2$ &  $0$ &  $2$&$+$ \\
       \hline
        $\zeta$ & {\bf 3} & $2/3$ & $7/3$ &  $(1,0,1)^T$&  $(2,2,3)^T$&$(+,+,-)^T$ \\
        $\xi$ & {\bf 3} & $2/3$ & $4/3$ &  $(1,0,1)^T$&  $(1,1,2)^T$&$(-,-,+)^T$ \\
\hline    
\end{tabular}
\caption{Lepton and scalar content with $a=1,2,3$ and $i=1,2$.}\label{tab}
\end{table}

Symmetry breaking takes place when scalar fields acquire vacuum expectation values (vevs), according to the pattern below, 
\bea\label{vevs}
\vev{\sigma}=\frac{v_\sigma}{\sqrt{2}},~\langle\chi\rangle=\frac{1}{\sqrt{2}}(0,0,w)^T,
\vev{\eta}=\frac{1}{\sqrt{2}}(v_1,0,0)^T,~ \langle\rho\rangle=\frac{1}{\sqrt{2}}(0,v_2,0)^T,~
\vev{\zeta}=\frac{1}{\sqrt{2}}(0,v_2',0)^T.
\eea
The spontaneous symmetry breaking process can be divided into two main steps
\bea 
SU(3)_c \otimes SU(3)_L \otimes U(1)_X \otimes U(1)_N &&\xrightarrow{v_\sigma, w} SU(3)_c \otimes SU(2)_L \otimes U(1)_Y \otimes M_P\\&&
\xrightarrow{v_1, v_2, v_2'}~SU(3)_c \otimes U(1)_Q \otimes M_P~,\nonumber
\eea 
for $v_\sigma, w\gg v_{EW}$ and $v_{EW} = (v_1^2 +v_2^2 + v_2'^2)^{1/2} = 246$ GeV.
After the breaking a residual matter-parity symmetry, $M_P$, is present 
\bea\label{MP}
    M_P &=& (-1)^{3(B-L) + 2s}~,
\eea
where $s$ is the particle's spin. As only the $M_P$-even scalar fields get vevs, $M_P$ remains as an absolutely conserved residual symmetry, in such a way that the lightest amongst the $M_P$-odd particles is stable, see Table \ref{tab}.

Taking into account the scalar fields in Table \ref{tab}, we can write down the most general renormalisable potential as
\bea\label{V}
V &=& \sum_{s} \left[\mu_s^2 (s^\dagger s)+\frac{\lambda_s}{2} (s^\dagger s)^2   \right] + \sum_{s_1,s_2}^{s_1>s_2} \left[ \lambda_{s_1 s_2 }(s_1^\dagger s_1)(s_2^\dagger s_2) \right] + \sum_{t_1,t_2}^{t_1>t_2} \left[ \lambda'_{t_1 t_2  }(t_1^\dagger t_2)(t_2^\dagger t_1)\right]\\
&&+ \frac{\mu_1}{\sqrt{2}} \eta\rho\chi + \frac{\mu_2}{\sqrt{2}} (\zeta^\dagger\rho)\sigma  + \lambda_1(\chi^\dagger \eta)(\zeta^\dagger \xi) +  \lambda_2  (\chi^\dagger \xi)(\zeta^\dagger \eta)  + \lambda_3 (\chi^\dagger \eta)(\xi^\dagger \rho)\nn\\
&&+ \lambda_4 (\chi^\dagger \rho)(\xi^\dagger \eta)
+ \lambda_5(\eta\zeta\chi)\sigma^* + h.c. ~,\nn
\eea 
where $s=\eta, \rho,\chi,\sigma,\zeta,\xi$ (all scalars), and $t=\eta, \rho,\chi,\zeta,\xi$ (scalar triplets only).
%

\subsection{Neutrino masses}  
\label{sec:neutrino-masses}


%
When it comes to the Yukawa interactions, we can write the Lagrangian below
\bea\label{Yuk}
- \mathcal{L}_{lep} &=& y^e_{ab}\,\overline{l_{aL}}\, \rho\, e_{bR} + y^N_{ab}\, \overline{l_{aL}}\, \chi N_{bR} + h_{ab}\, \overline{l_{aL}}\,(l_{bL})^c\, \xi^* +   \frac{(m_N)_{ab}}{2}\, \overline{(N_{aR})^c}\,N_{bR} + h.c.~, 
\eea
where $y^e, y^N, h$ and $m_N$ are complex $3\times3$ matrices, with $m_N$ being symmetric due to the Pauli principle. In contrast, $h$, the Yukawa coupling which governs the anti-symmetric contraction of three triplets, is necessarily anti-symmetric in family space.

Charged lepton masses come from the first term in Eq. (\ref{Yuk}) when $\rho$ acquires a vev: $M^e = y^e v_2/\sqrt{2}$. The neutral leptons $N_{iL}$ and $N_{iR}$ mix at tree-level when $\chi$ acquires a vev, and the mixing angle is defined by $\tan(2\theta_N)=2y^N w/m_N$.

\begin{figure}[htbp]
\begin{center}
\includegraphics[scale=.6]{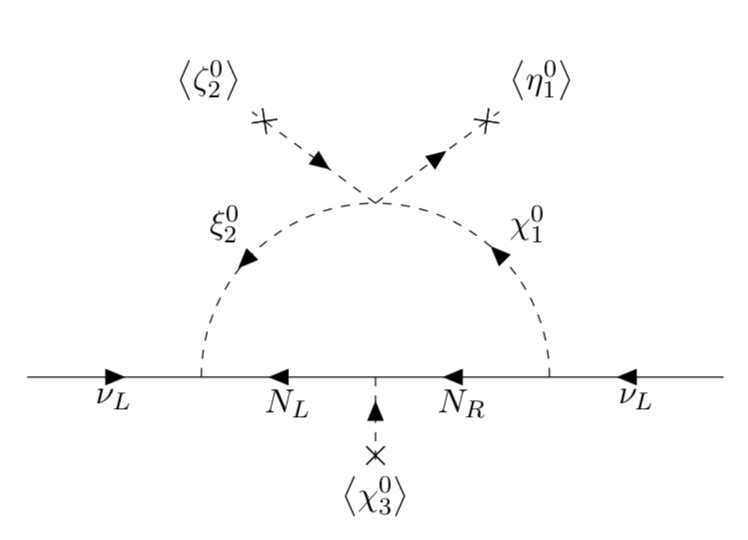}
\caption{1-loop ``scotogenic'' neutrino mass.}\label{fig1}
\end{center}
\end{figure}

Let us now turn our attention to the active neutrinos $\nu_L$. As neither the first component of
$\chi$ nor the second of $\xi$, both $M_P$-odd, acquires a vev, neutrinos are massless at tree-level. Nevertheless, neutrino masses are radiatively generated via the loop diagram in Fig.\ref{fig1}. 

In order to calculate this diagram, we go from the flavour basis where the internal scalar fields, $\chi_1^0$ and $\xi_2^0$, mix with each other as well as with $\eta_3^0$, to the physical basis. The physical fields are three CP-even and three CP-odd neutral scalars, $(S,A)_{1,2,3}$, which mix according to two mixing angles, $(\theta_{S,A})_{1,2}$, as shown in Ref. \cite{Leite:2019grf}. Thus, the one-loop neutrino masses can be written as{\small
\begin{align}
\label{eq:mnu}
m_{\nu}^{ab}&= \frac{h^{* ac} s_N c_N c_1}{8\pi^2} \left\{m_{N_1}\left[s_{S_2} c_{S_2}\left(Z\left(\frac{m_{S_1}^2}{m_{N_1}^2}\right)-Z\left(\frac{m_{S_2}^2}{m_{N_1}^2}\right)\right)-s_{A_2}c_{A_2}\left(Z\left(\frac{m_{A_1}^2}{m_{N_1}^2}\right)-Z\left(\frac{m_{A_2}^2}{m_{N_1}^2}\right)\right)\right]\right. \\
&\left.- m_{N_2}\left[s_{S_2} c_{S_2}\left(Z\left(\frac{m_{S_1}^2}{m_{N_2}^2}\right)-Z\left(\frac{m_{S_2}^2}{m_{N_2}^2}\right)\right)-s_{A_2} c_{A_2}\left(Z\left(\frac{m_{A_1}^2}{m_{N_2}^2}\right)-Z\left(\frac{m_{A_2}^2}{m_{N_2}^2}\right)\right)\right]\right\}_{cd} y^{N* db} + \left\{a\leftrightarrow b\right\}, \nn
\end{align}}
with $s_x\equiv \sin\theta_x,c_x\equiv \cos\theta_x$ and $Z(x)=\frac{x}{1-x}\text{ln}x$.
The antisymmetryic nature of the Yukawa matrix $h$ implies that the neutrino mass matrix is of rank two, leading to one massless neutrino.
This unique feature provides a novel origin for the masslessness of one neutrino to be contrasted with the usual models relying on missing partner mechanisms.

All fields running inside the neutrino mass loop are odd under the exactly conserved matter-parity symmetry. Therefore, the lightest among such particles, either a scalar or a fermion, is stable and can play the role of a WIMP dark matter.

\subsection{Neutrinoless double beta decay}

The model predicts Majorana neutrinos, which allows for the possibility of neutrinoless double beta ($0\nu\beta\beta$) decay to take place \cite{Schechter:1981bd}.
The dominant contribution to $0\nu\beta\beta$ decay is the standard one and depends on the effective Majorana mass, defined as
\be\label{mbb}
\vev{ m_{\beta\beta} } = |\cos^2 \theta_{12}\cos^2 \theta_{13} m_1+\sin^2 \theta_{12}\cos^2 \theta_{13} m_2 e^{2 i \phi_{12}} + \sin^2 \theta_{13} m_3 e^{2 i \phi_{13}}|~,
\ee 
where mixing angles and Majorana phases are neatly expressed in the symmetric parametrisation of the lepton mixing matrix~\cite{Schechter:1980gr}, and the neutrino masses $m_a$ obtained from Eq.~\ref{eq:mnu}. 

\begin{figure}[htbp]
\begin{center}
\includegraphics[scale=.15]{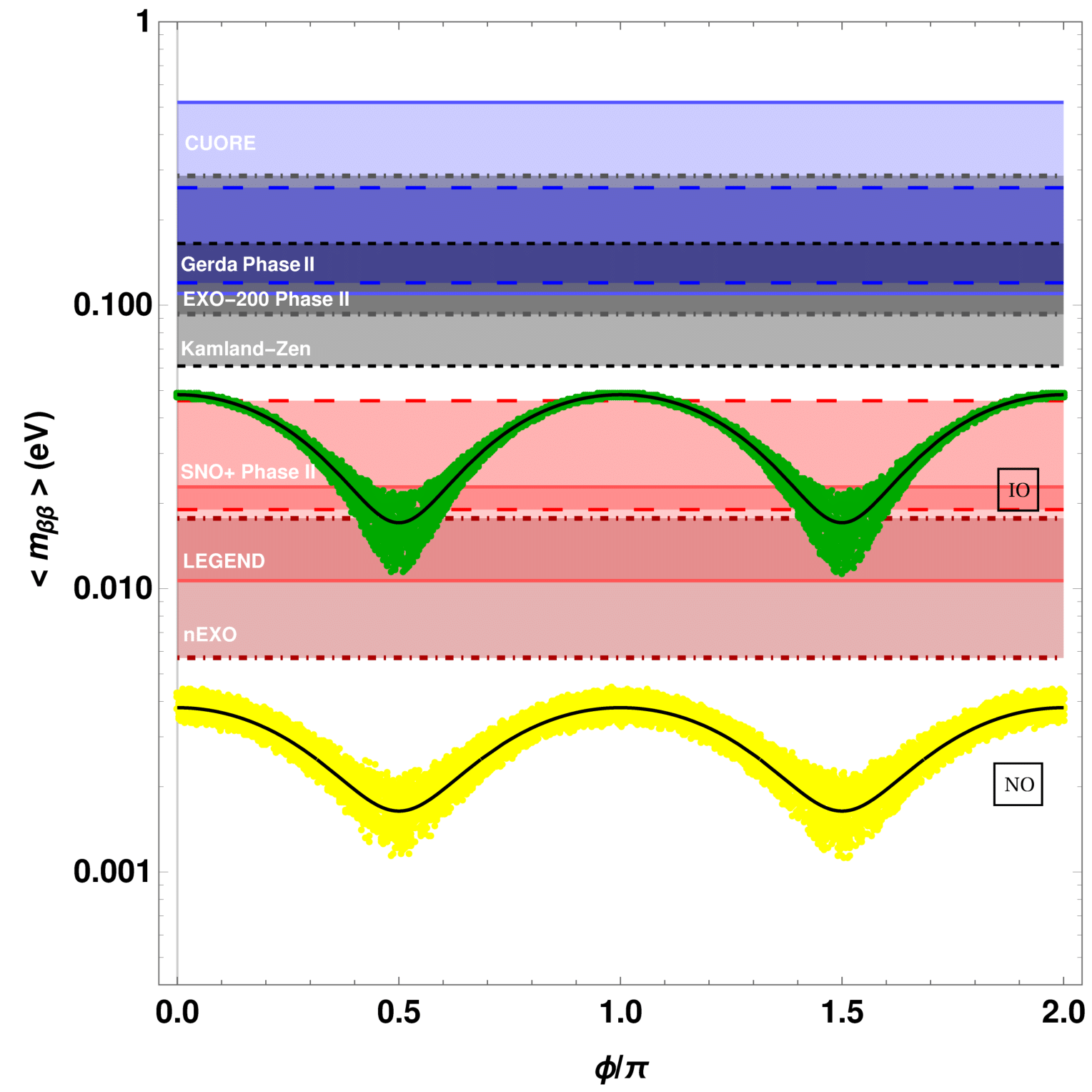}
\caption{Effective Majorana mass vs relative Majorana phase for the case of inverted (green) and normal (yellow) mass ordering.
Experimental limits and future sensitivities are displayed as horizontal bands.}\label{fig2}
\end{center}
\end{figure}

It is well-known that, in a generic model, this amplitude can vanish for normal-ordered neutrinos, currently preferred by oscillations~\cite{deSalas:2017kay}.
However, in our model, since one neutrino is massless, $\langle m_{\beta\beta} \rangle$ never vanishes, as shown in Fig. \ref{fig2}. 
\section{Conclusions}
\label{sec:Conclusions}

We have proposed an $\mathrm{SU(3)_c \otimes SU(3)_L \otimes U(1)_X \otimes U(1)_{N}}$ model where dark matter stability follows from the spontaneous breaking of the gauge symmetry.
Neutrino masses are generated at loop level and mediated by DM in a scotogenic fashion.
Our construction features a triplet scalar with anti-symmetric Yukawa couplings to neutrinos.
This leads to the prediction of a massless neutrino and a lower bound
for the $0\nu\beta\beta$ decay rate. 
Contrary to models where a massless neutrino arises from an {\it ad hoc} incomplete multiplet choice, 
here it is an unavoidable feature of the theory.

\section*{Acknowledgements}

\noindent

Work supported by the Spanish grants SEV-2014-0398 and FPA2017-85216-P
(AEI/FEDER, UE), PROMETEO/2018/165 (Generalitat Valenciana) and the
Spanish Red Consolider MultiDark
FPA2017-90566-REDC. J. L. acknowledges financial support under grant
2019/04195-7, S\~ao Paulo Research Foundation (FAPESP), while OP is
supported by the National Research Foundation of Korea, under Grants
No. 2017K1A3A7A09016430 and No. 2017R1A2B4006338.

\bibliographystyle{unsrt}
\bibliography{bibliography}

\end{document}